 \documentclass[final,5p,times,twocolumn]{elsarticle}

 \usepackage{graphicx}

\usepackage{amssymb}

\newcommand{\av}[1]{\langle{#1}\rangle}

\journal{Physics Letters B}

\begin{document}

\begin{frontmatter}



\title{Cosmic spherical void via coarse-graining and averaging non-spherical structures.}


\author[label1,label2]{Krzysztof Bolejko}
\ead{Krzysztof.Bolejko@astro.ox.ac.uk}
\author[label1]{Roberto A. Sussman}
\ead{sussman@nucleares.unam.mx}

\address[label1]{Instituto de Ciencias Nucleares, Universidad Nacional Aut\'onoma de M\'exico, A. P. 70--543, 04510 M\'exico D. F., M\'exico}
\address[label2]{Astrophysics Department, University of Oxford, 1 Keble Road, Oxford OX1 3RH, UK}


\begin{abstract}
Inhomogeneous cosmological models are able to fit cosmological observations without dark energy under the assumption that we live close to the ``center'' of a very large-scale under--dense region. Most studies fitting observations by means of inhomogeneities also assume spherical symmetry, and thus being at (or very near) the center may imply being located at a very special and unlikely observation point. We argue that such spherical voids should be treated {\it{only}} as a gross first approximation to configurations that follow from a suitable smoothing out of the non--spherical part of the inhomogeneities on angular scales. 
In this {\it Letter} we present a toy construction that supports the above statement. The 
construction uses parts of the Szekeres model, which is inhomogeneous and
anisotropic thus it also addresses the limitations of spherical inhomogeneities.
By using the thin-shell approximation (which means that the Israel-Darmois continuity conditions are not fulfilled between the shells)
we construct a model of evolving cosmic structures, containing several elongated supercluster-like structures with underdense regions between them,
which altogether provides a reasonable coarse-grained description of
cosmic structures.
While this configuration is not spherically symmetric, its proper volume average yields a spherical void profile of $250$ Mpc that roughly agrees with observations. Also, by considering a non-spherical inhomogeneity, the definition of a ``center'' location becomes more nuanced, and thus the constraints placed by fitting observations on our position with respect to this location become less restrictive.          
\end{abstract}


\end{frontmatter}


\section{Introduction}
Cosmological models allowing for non--trivial inhomogeneities have become a popular tool to 
analyze cosmological observations without the need of introducing an elusive dark energy source (for a review on a subject and explicit examples the reader is referred to Ref. \cite{Bbook}). Within this approach, the preferred configurations are Gpc-scale cosmic void models based on the spherically symmetric Lema\^\i tre-Tolman (LT) models \cite{L33,T34}, under the assumption that we live close to a center of a cosmic depression of radius around $1-3$ Gpc
\cite{AAG06,GH08,BW09}. These configurations are often criticized on the grounds that they violate the Copernican principle, since compliance with the cosmic microwave background (CMB) constraints implies that only one such Gpc structure is allowed and the observer cannot be further away from the origin than  $\sim 50$ Mpc \cite{AA07}
\footnote {The figure obtained in Ref. \cite{AA07} is 15 Mpc, 
but as shown in \cite{BW09} fluctuations of Hubble rate needed to fit observations
are $\Delta h \approx 0.08$ which means that for $d < 55$ Mpc the dipole is less than $3$ mK.}. However, as  suggested recently by Alexander et al. \cite{MVs} (see also \cite{BC10}), a void of radius $250$ Mpc is sufficient to explain the supernova observations, the power spectrum of the cosmic microwave background and is also consistent with Big Bang Nucleosynthesis, or Baryon Acoustic Oscillations.
For void structures of this size the Copernican Principle is not violated -- there can be many such structures as the upper size to violate the CMB constrains is 300 Mpc \cite{IS06,IS07}.
Also, restricting our position to be within 50 Mpc from the origin of a 250 Mpc void is a less  stringent constraint. However, one may argue that voids of radius $250$ Mpc are not observed in the filamentary structure characterizing our Local Universe.

In this {\it Letter} we address this issue by showing that these rather artificial spherical void structures need not exist in its pure form. Instead, they approximate configurations that can emerge after coarse-graining and averaging a sufficiently large scale region of a realistic lumpy Universe in which the density distribution is far from spherical. We consider for this purpose a non--spherical inhomogeneous and anisotropic Szekeres model, prescribing its free functions by means of a thin-shell approximation \cite{IS06,IS07,TV87,K00}, leading to a reasonable coarse-grained description of realistic cosmic structures. Initial conditions are  defined at the last scattering surfaces to show that the model can evolve from small early universe initial fluctuations, hence we achieve consistency with models of structure formation.
We show that averaging this inhomogeneous non-spherical configuration leads to 
a cosmic void that is qualitatively similar to the spherical models discussed by Alexander et al. \cite{MVs}.

It is important to emphasize that the lack of spherical symmetry in the Szekeres model removes the unique invariant nature of the center location of models with this symmetry. This is a relevant feature of the model, as our being sufficiently near this center is a strong constraint that the fitting of observations place on spherical LT models. As we argue further below, this constraint may be less restrictive in a non-spherical Szekeres model.

\section{Szekeres model}

\subsection{Einstein's equations}

The metric of the Szekeres model takes the following form \cite{S75}

\begin{equation}
{\rm d} s^2 =   {\rm d} t^2 - \frac{(\Phi' - \Phi {  {\cal E}}'/ {  {\cal E}})^2}
{\epsilon - k} {\rm d} r^2 - \frac{\Phi^2}{{\cal E}}({\rm d} x^2 + {\rm d} y^2),
\label{ds2}
 \end{equation}
where $\Phi=\Phi(t,r)$, {the prime denotes the partial derivative with respect to $r$,
$\Phi' = \partial \Phi / \partial r$},  and
\begin{equation}
{  {\cal E}} = \frac{S}{2} \left[ \left( \frac{x-P}{S} \right)^2
+ \left( \frac{y-Q}{S} \right)^2 +\epsilon \right],
\label{Edef}
\end{equation}
while $k(r), S(r), P(r), Q(r)$ are arbitrary functions;
 $\epsilon$ is a constant: the $\epsilon = -1$ case is often called the  
quasi-hyperbolic Szekeres model,
$\epsilon = 0$ quasi-plane,
and $\epsilon = 1$ quasi-spherical (for a detailed discussion on these models see \cite{HK08,K08,HK02}) -- 
in this {\it Letter} we only consider the quasispherical case.
{The coordinate system in which the metric has the form (\ref{ds2}) can be interpreted as a stereographic projection of polar coordinates \cite{HK02,PK06}.
For the quasispherical case the transformation is of the following form:}

 \[ \{ x - P, y-Q \} = \left\{ S {\rm cot} \left( \frac{ \theta }{2} \right) \cos (\phi), S {\rm cot} \left( \frac{ \theta }{2} \right) \sin (\phi) \right\} \] 
{  
Then, using the $(\theta$, $\phi)$ coordinates we can rewrite ${\cal E}'/   {\cal E}$ as \cite{HK02,PK06}
\begin{equation}
\frac{{\cal E}'}{{\cal E}} = -\frac{S,_r \cos \theta + \sin \theta \left( P,_r \cos \phi + Q,_r
\sin \phi \right)}{S}. \label{nuz}
\end{equation}
However, under the above transformation the metric becomes non-diagonal \cite{Bsz1,Bsz2}.
Thus, for some  applications it is more convenient to use the $(t,r,x,y)$ coordinates -- see Sec. \ref{Ave}. Notice that we use the $(t,r,\theta,\phi)$ coordinates in Sec. \ref{Pos}.
}

Einstein's equations for a dust source associated with (\ref{ds2})--(\ref{Edef}) reduce to
\begin{equation}\label{vel}
\dot{\Phi}^2 = - k(r) + \frac {2 M(r)} {\Phi},
\end{equation}
where $M(r)$, is an arbitrary function related to the density $\rho$ via:
\begin{equation}
\kappa \rho = \frac{2M' - 6 M {\cal E}'/{\cal E}}{\Phi^2 ( \Phi' - \Phi {\cal E}'/{\cal E})},
\label{rho}
\end{equation}
where $\kappa = 8 \pi G$. Whenever $\Phi' = \Phi {\cal E}'/{\cal E}$ and $ M' \ne 3 M {\cal E}'/{\cal E}$ hold
a shell crossing singularity occurs, which (in 
a quasi-spherical model) may occur along a circle, or, in
exceptional cases, at a single point, and not at a whole surface in the $(t,r)$
plane, as is the case in LT models \cite{HK02,PK06}.

As in a spherical LT model, the bang time function follows from (\ref{vel})
\begin{equation}\label{tbf}
\int\limits_0^{\Phi}\frac{{\rm d} \widetilde{\Phi}}{\sqrt{- k + 2M /
\widetilde{\Phi}}} = t - t_B(r).
\end{equation}

\subsection{Set-up}

While the previous equations indicate that the Szekeres model is specified by 6 functions, choosing appropriate coordinates eliminates one of the independent functions. Thus, we must provide 5 functions as initial conditions to specify the model. In particular, we will specify  the  functions $S,P,Q,t_B$, and $M$.

In order to achieve with a Szekeres model the most realistic possible description of cosmic structures and structure formation, we specify below suitable forms for these 5 functions and define our model at the last scattering instant, so that it evolves from small initial fluctuations up to the present cosmic time.
The algorithm that we use in the calculations can be defined as follows:

\begin{enumerate}
 \item
The radial coordinate is chosen to be the areal radius at the last scattering instant
$\bar r = \Phi(r, t_{i})$. However, to simplify the notation we will omit the bar and denote the new radial coordinate by $r$.

\item
The chosen asymptotic cosmic background is an 
open Friedman model\footnote{{  The standard conventional analysis of cosmological 
observations appears to imply a spatially flat background. However, when the assumption of homogeneity is relaxed spatial flatness is no longer required.
For specific examples and a comprehensive discussion on this issue the reader is kindly referred to 
\cite{BW09,ClRe2010}.}},
i.e. $\Omega_m = 0.3$ and $\Lambda=0$. 
The background density is then given by
   
 \begin{equation}
   \rho_b = \Omega_m \times \rho_{cr} = 0.3 \times \frac{3H_0^2}{8 \pi G} (1+z)^3,
 \end{equation}
where the Hubble constant is $H_0 =70$ km s$^{-1}$ Mpc$^{-1}$.

\item   The age of the universe is given by (\ref{tbf}).
Notice that inhomogeneities may affect the age relation via the bang time function $t_B$.
However, since we define our model at the last scattering when the Universe is expected to be very close to homogeneity, we assume that $t_B =0$. As a consequence, the age of the Universe is everywhere the same (as in the homogeneous background Friedmann model) and is equal to $t_i = 471,509.5$ years\footnote{The age of the universe at the last scattering instant is below 400, 000 yr. 
This is due to the presence of radiation that is not negligible at the decoupling instant. However, the Szekeres model only describes the pressureless configuration (dust).
That is why when calculating the initial instant we neglect the pressure and only consider dust. If radiation is neglected then the age of the universe at the decoupling instant is larger than 400, 000 yr.} (see \cite{P80} for details).

\item The function $M(r)$ is given by

\[ M(r) = 4 \pi \frac{G}{c^2} \int_0^r 
 \rho_b(1+ \delta\bar{\rho})\, \bar r^2\,{\rm d} \bar r,\]
where $\delta \bar{\rho}=-0.005{\rm e}^{-(\ell/100)^2}+ 0.0008 {\rm e}^{-[(\ell-50)/35]^2}
+0.0005 {\rm e}^{-[(\ell-115)/60]^2}	
+ 0.0002 {\rm e}^{-[(\ell-140)/55]^2}$,
and $\ell \equiv r/$ 1 kpc.

\item The function $k(r)$ can be calculated from (\ref{tbf}).

\item
The functions $Q, P,$ and $S$ are defined as follows

$S=1 \Rightarrow S' = 0,$

${\cal D} = 1.05 (1+r)^{-0.99} {\rm e}^{-0.004 r},$

$Q' = {\cal D}$, $P'=0$ for $\ell \leqslant 27,$

$Q' = -{\cal D}$, $P'=0$ for $27 < \ell \leqslant 35,$

$Q' = 0$, $P'=- {\cal D}$ for $35 < \ell \leqslant 41,$

$Q' = 0$, $P'= {\cal D}$ for $41 < \ell \leqslant 51.5,$

$Q' = 0.88 {\cal D} $, $P'=- 0.5 {\cal D}$ for $51.5 < \ell \leqslant 61,$

$Q' = 0.71 {\cal D} $, $P'=0.71 {\cal D}$ for $61 < \ell \leqslant 69,$

$Q' = 0 $, $P'=- {\cal D}$ for $69 < \ell \leqslant 77,$

$Q' = - {\cal D} $, $P'= 0$ for $77 < \ell \leqslant 86.5,$

$Q' = 0.74 {\cal D} $, $P'=-0.74 {\cal D}$ for $86.5 < \ell \leqslant 96,$

$Q' ={\cal D} $, $P'= {\cal D}$ for $96 < \ell \leqslant 102,$

$Q' = - {\cal D} $, $P'= 0$ for $102 < \ell \leqslant 115,$

$Q' = {\cal D} $, $P'= 0$ for $115 < \ell \leqslant 129,$

$Q' = 0 $, $P'=- {\cal D}$ for $\ell >129.$

\item Once the model is specified, its evolution is calculated from 
eq. (\ref{vel}).

\item Density distribution at the current instant is then evaluated
from (\ref{rho}). 

\end{enumerate}

As can be seen from the expressions above, some of the functions are not continuous, hence {  the Israel-Darmois matching conditions for spacetimes are not fulfilled. Therefore,}
this model should be regarded as a thin-shell approximation. This approximation, in its original form \cite{IS06,IS07,TV87,K00}, consists of: (i) two Friedmann regions 
and (ii) transitions between them modeled by thin shells with negligible effects.
Here we apply a similar approach, but instead of Friedmann regions
we consider Szekeres regions. This approach is justified by the fact
that in the Szekeres models each surface of constant $r$ (a 2--sphere in our case)
evolves as an independent Friedmann model -- as seen from eq. (\ref{vel})
each shell can be considered as a Friedmann model whose evolution
is given by its Friedmann equation (\ref{vel}), a different Friedmann equation for each sphere.
The only restriction is that the model must be free from shell crossings  -- an inner sphere cannot expand faster
than an outer sphere. This has been taken into account and
the above defined model evolves without shell crossings from the initial
instant (last scattering) until the present moment.

\begin{figure}
\begin{center}
\includegraphics[scale=0.24]{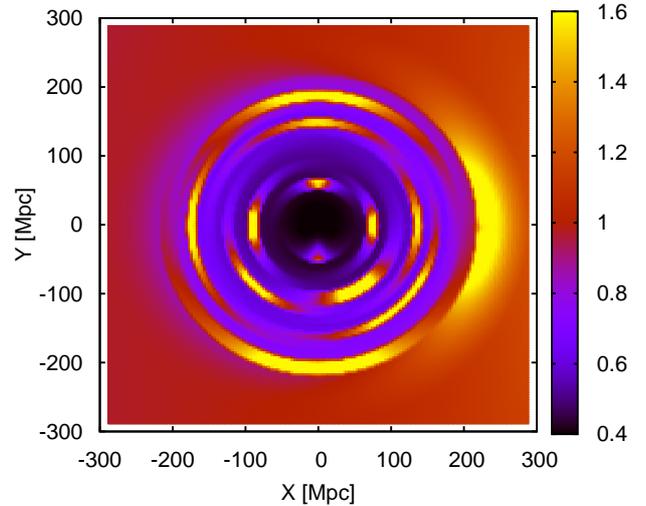}
\caption{{  The present-day color-coded density distribution $\rho/\rho_0$ (where $\rho_0$ is density
of the homogeneous background model). Brighter colors indicate a high-density region, darker low-density region.
The presented surface is a horizontal cross-section, i.e. $\theta = \pi/2$. For a discussion on how realistic this structure is see Sec. \ref{How}.}}
 \label{m2d}
\end{center}
\end{figure}

The density distribution for our model is depicted in Fig. \ref{m2d} {  (we remark that this model does not admit Killing vectors)}. We are using 
intuitive Cartesian coordinates (see Ref. \cite{Bsz1,Bsz2} for the corresponding 
transformation and description of these coordinates).
It should now be clear why we have selected the above set of functions
to define the Szekeres model. As seen in Fig. \ref{m2d}, we  have all together 13 different sets with  the functions $\{M,t_b,Q,P,S\}$ and so we have 13 different overdensities.
By changing the form of the above functions we can change the position, size, and the amplitude of the overdensities (see Ref. \cite{Bsz1,Bsz2} for a detailed discussion).
It is important to remark that the very special choice of free functions is not an act of ``fine tuning'' (in the sense of a tricky manipulation of parameters), but an effort to approximate the known Cosmography as best as possible. Hence, if new data would modify or improve the known density distribution it would be straightforward to modify these functions to approximate this data, which would then lead to a structure that is different from that presented in Fig. \ref{m2d}.

\subsection{How realistic this model is ?}\label{How}

As shown in Fig. \ref{m2d}, the model under consideration contains structures such as voids
and elongated supercluster--like overdensities.
It has large overdensities around $\sim 200$ Mpc 
({  on the left in Fig. \ref{m2d}}) that compensate 
the underdense regions and allow the model to be practically homogeneous at $r>300$ Mpc. 
In the real universe we observe very massive matter concentrations -- the 
Shapley Concentration roughly at the distance of 200 Mpc, or 
the Great Sloan Wall at the distance of 250-300 Mpc. In the opposite
direction on the sky we find the Pisces--Cetus and Horologium--Reticulum,
which are massive matter concentrations located at a similar distance. 
We refer the reader to Fig. 44 of Ref. \cite{SW}, which provides a density map of the Local Universe reconstructed from the 2dF Galaxy Redshift Survey Survey using
Delaunay Tessellation Field Estimator\footnote{This figure is
also available at http://en.wikipedia.org/wiki/File:2dfdtfe.gif}. 
Also, the inner void seen in Fig. \ref{m2d} is consistent with what is observed in the Local Universe -- it appears that our Local Group 
is not located in a very dense region of the Universe, rather it is located in a less dense region surrounded by large overdensities like the Great Attractor
on one side and the Perseus--Piscis supercluster on the other side. Both are located at around 50 Mpc --- see Fig. 19 of \cite{LU} that provides the density reconstruction of the Local Universe using the POTENT analysis.

While still far from a perfect ``realistic'' description, the model displayed in Fig. \ref{m2d} exhibits
the main features of our local Universe. It should be therefore treated as a ``coarse-grained'' approximation to study local cosmic dynamics by means of a suitable exact solution
of Einstein's equations. Such approximation is, evidently, far less idealized than the gross one that follows from spherically symmetric LT models.

\subsection{Position of the ``center''}\label{Pos}

Since the model we are considering is not spherically symmetric, there is no invariant and unique characterization of a center worldline. Instead, for every 2--sphere corresponding to a fixed value of $r$, we have (at least) two locations that can be considered appropriate generalizations of the spherically symmetric center: the worldline marked by $r=0$ where the shear tensor vanishes, which defines a locally isotropic observer (cf. eq (16.29) of Ref. \cite{PK06}), and the ``geometric'' center of the 2--sphere whose surface area is  $4\pi \Phi^2$.
  
\begin{figure}
\begin{center}
\includegraphics[scale=0.7]{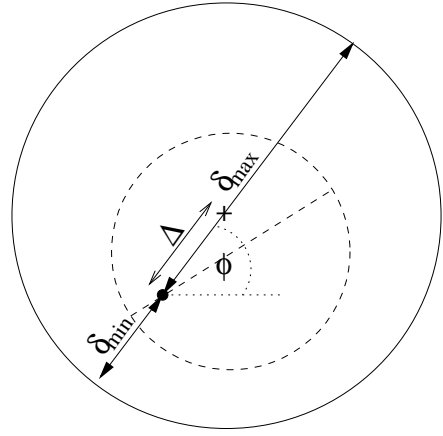}
\caption{{  Schematic representation of the quasispherical Szekeres model -- different surfaces of constant $t$ and $r$ (spheres)
have different centers, which do not coincide with the origin marked by $r=0$ (denoted by a black dot).
The center of the larger sphere (geometric center) is depicted by a cross. 
The maximal distance from the origin to the surface of the sphere is $\delta_{max}$ and the minimal is $\delta_{min}$. The distance between
the geometric center and the origin is denoted by $\Delta$.}}
 \label{fg2}
\end{center}
\end{figure}

\begin{figure}
\begin{center}
\includegraphics[scale=0.7]{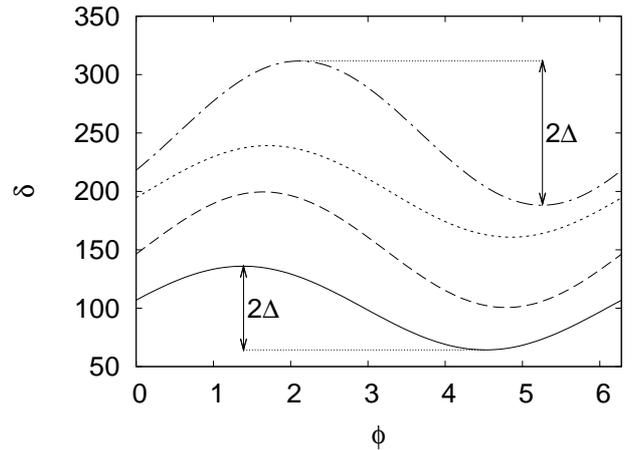}
\caption{{  Proper distance, $\delta$, given by (\ref{PropDis}) as a function of $\phi$ for 4 different cases:
$\Phi(t_0,r) = 250$ Mpc (the dash-dotted line), $\Phi(t_0,r) = 200$ Mpc (the dotted line), $\Phi(t_0,r) = 150$ Mpc (the dashed line),
$\Phi(t_0,r) = 100$ Mpc (the solid line). For all cases $\theta = \pi/2$.}}
 \label{fg3}
\end{center}
\end{figure}

Since the 2--spheres of constant $r$ in a quasi-spherical Szekeres model are non-concentric, their
geometric center does not coincide with the
the point of vanishing shear $r=0$ 
{  -- this is schematically presented in Fig. \ref{fg2}. 
As a result, the distance from the origin to the surface of the sphere depends on the direction marked by the angles $(\theta,\phi)$:}
\begin{equation}
\label{PropDis}
\delta(r,\theta,\phi) =  \int_0^r {\rm d} \tilde{r} \frac{\Phi' - \Phi {  {\cal E}}'/ {  {\cal E}}}
{\sqrt{1- k}} , 
\end{equation}
{  and thus, the displacement $\Delta$ between 
the origin and the geometric center of a sphere marked by comoving radius $r$ is given by}
\[\Delta = \frac{\delta_{{\rm max}} - \delta_{{\rm min}}}{2}, \]
where $\delta_{{\rm max}} = {\rm max} (\delta)$,
$\delta_{{\rm min}} = {\rm min} (\delta)$.
{  
As seen from (\ref{nuz}), the maximal and minimal value of ${\cal E}'/   {\cal E}$ for our model (where $S'=0$) corresponds to $\theta = \pi/2$.
The distance, $\delta$, as a function of $\phi$ 
for 4 cases [(1) present-day areal radius is $\Phi(t_0,r) = 100$ Mpc,
(2)  $\Phi(t_0,r) = 150$ Mpc,  (3) $\Phi(t_0,r) = 200$ Mpc, and
(4)  $\Phi(t_0,r) = 250$ Mpc] is depicted by Fig. \ref{fg3}.}
As seen, for example, for a sphere whose present-day area radius is $\Phi=100$ Mpc the model under consideration yields   
a displacement of  $\Delta =  36$ Mpc
towards $\phi \approx 80^\circ$ direction.
While for $\Phi = 250$ Mpc we have $\Delta =  62$ Mpc towards
$\varphi \approx 120^\circ$.

Different 2--spheres of constant $r$ have their centers displaced (with respect to the
coordinate origin where 
the shear vanishes) with different values and towards
different directions. Hence, there is no uniquely defined 
geometric center -- a center of one sphere does not coincide with a center
of another sphere and moreover it also does not coincide with the
point where  shear vanishes. Since the center in spherical LT models is also a locally isotropic observer (where shear vanishes), and fitting observations in these models restricts our cosmic location to be within a given maximal separation from this observer, then it is reasonable to expect that similar restrictions should emerge in a Szekeres model given in terms of maximal separation from the local isotropic observer at the coordinate origin where shear vanishes. Therefore, the displacement of the geometric center from this origin would make our location in a Szekeres model less special and improbable than in spherically symmetric models where both center locations (local isotropic observer and geometric center) necessarily coincide.

\section{Averaging}\label{Ave}

As shown in Ref. \cite{Bav}, the proper 3--dimensional volume in space slices orthogonal to the 4--velocity ($t=$ constant) in a Szekeres model is

\begin{eqnarray}
&& V_{\mathcal{D}} =  \int\limits\limits_{0}^{r_{\mathcal{D}}} {\rm d} {r} \int\limits_{-\infty}^{\infty} {\rm d} x
\int\limits_{-\infty}^{\infty} {\rm d} y \sqrt{-g} \nonumber \\
&& = 4 \pi \int\limits_{0}^{r_{\mathcal{D}}} {\rm d} {r}
\frac{\Phi^2 \Phi'}{\sqrt{1-k}}  \equiv 4 \pi R_{\mathcal{D}},
\end{eqnarray}
and thus, the proper volume averaged density is spherically symmetric ({\it{i.e.}}  independent of $x$ and $y$), even if the density itself is far from a spherical distribution:

\begin{eqnarray}
&& \av{ \rho}(r_{\mathcal{D}}) = \frac{1}{V_{\mathcal{D}}} 
\int\limits\limits_{0}^{r_{\mathcal{D}}} {\rm d} {r} \int\limits_{-\infty}^{\infty} {\rm d} x
\int\limits_{-\infty}^{\infty} {\rm d} y \sqrt{-g} ~  \rho = \nonumber \\
&& \frac{1}{\kappa R_{\mathcal{D}}} \int\limits_{0}^{r_{\mathcal{D}}}{\rm d}  {r} \frac{2M'}{\sqrt{1 - k}}. 
\label{arho}
\end{eqnarray}

\begin{figure}
\begin{center}
\includegraphics[scale=0.24]{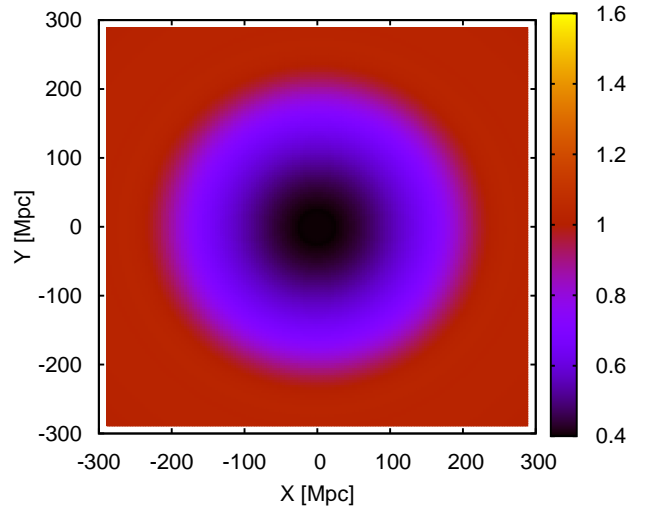}
\caption{Volume average density distribution  $\av{\rho}(r_{\mathcal{D}}) / \rho_0$
(where $\rho_0$ is the background density). 
{  Brighter colors indicate a high-density region, darker low-density region.
The displayed surface is a horizontal cross-section, i.e. $\theta = \pi/2$.}
As follows from (\ref{arho}), all ``angular'' dependence of inhomogeneities has been smoothed out.}
 \label{avd}
\end{center}
\end{figure}

\begin{figure}
\begin{center}
\includegraphics[scale=0.67]{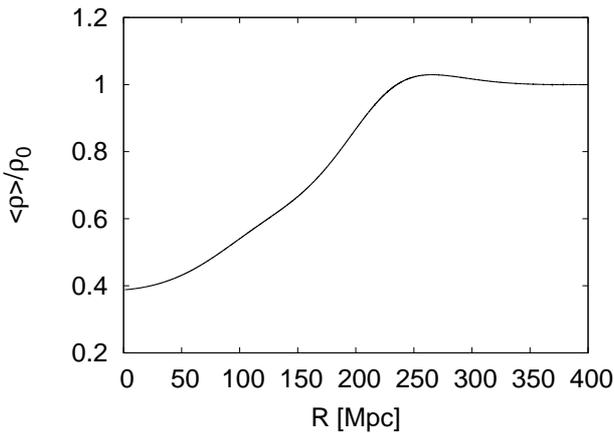}
\caption{Density profile after averaging $\av{\rho}(r_{\mathcal{D}}) / \rho_0$
(where $\rho_0$ is the background density).
Since  $\av{\rho}(r_{\mathcal{D}})$ in (\ref{arho}) is independent of $x$ and $y$, its dependence on $r$ has the form of  a cosmic void whose radius is approximately $250$ Mpc (analogous to the spherical voids of reference \cite{MVs}).}
 \label{avp}
\end{center}
\end{figure}

\noindent
This spherical volume--averaged density distribution,  $\av{\rho}(r_{\mathcal{D}})$, evaluated as a function of $r_{\mathcal{D}}$, is displayed by Fig. \ref{avd}, while its profile along the radial direction is presented in Fig. \ref{avp}.  As can be seen from these figures, the ``angular'' ({\it{i.e.}} $x,y$) dependence of an inhomogeneity (which is a highly non--spherical coarse grained density distribution) has been smoothed out by the averaging process, resulting into an averaged distribution that is equivalent to a spherical cosmic void whose radius is approximately $250$ Mpc (as in Ref. \cite{MVs}).

It is important to remark (and clarify) several issues. First, the field source is the local non--spherical density, $\rho$, not its spherically symmetric average, $\av{ \rho}$. Hence, the former (and not the latter) must be considered in the study of the general relativistic dynamics of the inhomogeneity, as well as in all computations related to observations (based on null geodesics). Second,  considering the dynamics of $\av{ \rho}$ by means of the suggested averaging procedures for General Relativity \cite{averages} is beyond the scope of this {\em Letter}. Instead, since $\av{ \rho}$ follows from a proper volume averaging procedure that is consistent for covariant scalars, we simply regard this averaged density as an approximation that conveys useful non--local information on $\rho$, even if it cannot account for the latter's description of the intricacies of local detail. 

While the proper volume average $\av{ \rho}$ in quasi--spherical Szekeres is necessarily a spherical distribution \cite{Bav}, it is not obvious (and not possible to know beforehand) the form of its radial density profile, though some sort of void profile should be expected given the observed void dominance in cosmic structure. In this context, we regard as an important result the fact that the simple shallow void profile displayed in figure \ref{avp} emerged by averaging the specific non--spherical $\rho$ that we proposed, which, as mentioned before, provides a reasonable course grained description of realistic cosmic structure. The fact that the void profile of figure \ref{avp} (and not any other more complicated void--like or clump/void mixed profile) resulted is an encouraging signal, as it provides a theoretical connection with spherical void models (as those studied in \cite{MVs} whose profiles resemble that of figure \ref{avp}). We also have a concrete example suggesting that spherical models, as idealized approximations, may be analogous to the average (first integral momentum approximation) of well motivated (and less idealized) non--spherical models.  As a consequence, the use of a Szekeres model seems to suggest that results obtained by means of spherical LT models may be robust: while local non--spherical information could still provide important refinements, and is needed for computations involving null geodesics (specially when fitting CMB constraints), it is likely that basic bottom line information is already contained in the spherical voids constructed with LT models.

\section{Conclusions}

We have presented in this {\it Letter} an explicit example of how
a non-spherical 
construction using parts of an exact solution of Einstein's equations can provide a reasonable coarse-grained description of a present day
cosmic structure that has evolved from small initial fluctuations
at the last scattering surface. This coarse grinned density distribution produces (after averaging) a cosmic void profile that is analogous to density profiles in spherical voids of radius $\sim 250$ Mpc. As mentioned before, a void of at least this size is required to explain cosmological observations without the need for dark energy.

We emphasize that we do not claim that the coarse-grained description given by our model is completely realistic, though it contains essential features and thus represents a significant improvement over the description of observed cosmic structure in spherical LT models used in the literature.
Our model uses the thin shell approximation which means that the 
 Israel-Darmois conditions are not fulfilled between the shells.
 Also, the model becomes practically homogeneous at a distance of $\sim 350$ Mpc, and thus it only describes one small region of our Universe, which should be composed of many other similarly sized structures. Such a construction would be consistent with 
supernova observations and the power spectrum of the cosmic microwave background, and 
would not violate the Copernican Principle and the constrains form the Rees-Sciama
effect -- to violate the R--S effect every single component structure must be larger than $\sim 300$ Mpc
\cite{IS06,IS07}.

The model we have presented is one among the first attempts in using the Szekeres solution as a theoretical and empiric tool to study and interpret cosmological observations \cite{MI1,MI2,BC10,KB10}. This opens new possibilities for inhomogeneous cosmologies, as this is the most general available cosmological exact inhomogeneous and anisotropic solution of Einstein's equations. The model provides a more nuanced and much less restrictive description of the need to constrain our location with respect to a center location. It is also a concrete example that illustrates the possibility that a mildly increasing void profile (required by observations) can emerge if local structures are coarse-grained and then averaged.  
Of course, notwithstanding these appealing features, the model and its assumptions must be subjected to hard testing by data from the galaxy redshift surveys, and evidently the more comprehensive this data can be the better it can be used for this purpose. 
Unfortunately current surveys like 2dF of SDSS do not cover the whole sky
and only focus on small angular regions of it.
However in the near future this limitation may be overcome -- for example, Sky Mapper\footnote{http://msowww.anu.edu.au/skymapper/} aims to cover
the whole southern sky which will provide sufficient data to test possibilities suggested and elaborated in this {\it Letter}. A more 
comprehensive and detailed article on the model proposed here is currently under elaboration and will be submitted soon for publication.

\section*{Acknowledgments}
We acknowledge financial support from grant PAPIIT-DGAPA IN119309. KB is very grateful to RAS  and ICN-UNAM, where most of this research was carried out, for their hospitality. The Marie Curie Fellowship (PIEF-GA-2009-252950) is also acknowledged.


\end{document}